\documentclass[12pt,english,showpacs,floatfix,reprint,amsmath,amssymb,prb,superscriptaddress]{revtex4-1}

\usepackage{graphicx}
\usepackage{graphics}
\usepackage{amsmath}
\usepackage{amsfonts}
\usepackage{amssymb}
\usepackage{epstopdf}
\usepackage{makeidx}
\usepackage{epsfig}
\usepackage{bm}
\usepackage[colorlinks=true,citecolor=blue,linkcolor=blue,linktocpage=true,pagebackref=false]{hyperref}
\usepackage{color,soul} 
\usepackage{float}

\newcommand{\be}{\begin{equation}}
\newcommand{\ee}{  \end{equation}}
\newcommand{\ba}{\begin{eqnarray}}
\newcommand{\ea}{  \end{eqnarray}}

\begin{document}

\title{Analysis of a networked social algorithm for collective selection of %voting 
a committee of representatives}

\author{Alexis R. Hern\'andez}
\affiliation{Instituto de F\'{\i}sica,
Universidade Federal do Rio de Janeiro, Rio de Janeiro, Brazil}
\author{Carlos Gracia-L\'azaro}
\affiliation{Institute for Biocomputation and Physics of Complex Systems (BIFI), Universidad de Zaragoza, Zaragoza, Spain}
\author{Edgardo Brigatti}
\affiliation{Instituto de F\'{\i}sica,
Universidade Federal do Rio de Janeiro, Rio de Janeiro, Brazil}
\author{Yamir Moreno}
\affiliation{Institute for Biocomputation and Physics of Complex Systems (BIFI), Universidad de Zaragoza, Zaragoza, Spain}
\affiliation{Department of Theoretical Physics, Faculty of Sciences, Universidad de Zaragoza, Zaragoza, Spain}
\affiliation{ISI Foundation, Turin, Italy} 
\date{\today}

\begin{abstract}  
A recent work ( Hern\'andez {\it et al.} \cite{nos}) introduced a networked voting rule 
supported by a trust-based social network, where indications of possible 
representatives were based on individuals opinions. 
Individual contributions went beyond a simple vote-counting and were
based on proxy voting. These mechanisms generated a high level of representativeness of the selected committee,
weakening the possibility of relations of patronage.
By incorporating the integrity of individuals and its perception, here we address the question of the trustability of the resulting committee. Our results show that this voting rule provides high representativeness for small committees with a high level of integrity. Furthermore, the voting system displays robustness to strategic and untruthful application of the voting algorithm.
\end{abstract}

\maketitle

\section{Introduction}

The form of the participation of common people to contemporary and 
complex democracies is a central issue in the social debate.
Many transformation and possible innovation have been recently discussed \cite{ColemanStephen,Chadwick,Hindman}, 
often forced by  the widespread use of digital technologies,
which redesign our interactions in politics and society.
A general problem, which ranges from national to neighborhood scales, 
is the problem of selecting an exemplary group of representatives to make decisions on behalf of the community
\cite{condorcet,classical,katz}.

Despite that the theoretical and philosophical debate over these issues has been prolific \cite{Noveck,Shane}, examples
of empirical construction of new algorithms have been relatively limited \cite{rodriguez,yamakawa,boldi,nos}. 
Recently, Hernandez {\it et al.} introduced a new social algorithm for collective selection of a committee of representatives \cite{nos}.
This algorithm for collective selection is developed starting from a standard situation where each voter is allowed to vote for only one candidate. However, the elected representatives are the ones who obtain a better rank among their counterparts, in a way that individual contributions go far beyond a simple vote-counting.

The introduced formal algorithm presents new specific features which could improve legitimation and fairness
of governance. The lists of candidates are not fixed in advance, but they emerge as a self-organized process controlled by the voting rules. This fact introduces an effective participation and engagement of the whole community, in contrast to top-down rigid lists of candidates. The voters express not preferences, but opinions, which determine their indications about whom they would like to see as their representatives. Finally, the new proposed mechanism improves the  representativeness  of the committee,
weakening the possibility of relations of patronage and clientelism. Additionally, the mechanism of votes aggregation is supported by a self-declared confidence circle, which defines a network of trusted individuals. This trust-based social network, which can be implemented on an online platform, is a fundamental ingredient that allows for direct accountability of the elected committee. 
Even if based on a local network, it can naturally scale to national sizes, translating to those larger scales an efficacious accountability typical of small-sized communities. 

In this work, we analyze a new aspect that can be introduced in the original algorithm. Specifically, we incorporate the possibility of a form of direct choice of individuals over the possible elected representatives. By doing that, we mitigate the aspect that voters determine their indications about whom they would like to see as their representative through opinions, valuing the principle that individuals directly select candidates. This new ingredient is implemented by introducing the expression of a preference among the contact network of individuals. Preferences act as a weight on the original opinion-based ranking algorithm in such a way that higher rates for these preferences are assigned to individuals considered more apt to participate in the committee.

The previous mechanism improves the legitimation, fairness, and effectiveness of the committee. In fact, overlaps, which are not controlled by voters, are weighted by a term subjectively assigned by the individuals. This weight should encourage a check on incompetence and corruption. Incompetence because an equal say for every individual is not necessarily always desired. Corruption as the preference should be proportional to the person who demonstrates and promises true integrity: sound ethical principles and trust. As each voter knows their representatives and each committee member knows to whom he is accountable, this fact will allow a strong control over representatives' actions.

The purpose of this work is to present and characterize in depth the new social algorithm throughout computational analyses. In Sec. II we describe the details of the algorithm. Sec. III is devoted to test the new voting rule, modeling the behavior of the selected committee. The quality of the elected committee is assessed looking at how much their final decisions are consistent with the personal opinions of the community and estimating the general integrity of the elected committee. Finally, Sec. IV presents some discussions of our results and concluding remarks.

%%%%%%%%%%%%%%%%%%%%%%%%%%%%%%%%%%
\section{The model}

Let us assume a system composed by $N_e$ electors interacting on an internet-based platform.
The platform allows the voters to declare who belongs to their interaction circle,  which renders a network of well-known individuals.
Voters also declare their perception of integrity for each individual $k$ belonging to their 
interaction circle. This perception is condensed in a scalar value $I_{jk} \in [0,1]$, which represents the perception that individual $j$ has about the integrity of individual $k$. 

In a following step, voters manifest their opinions on $N_i$ issues. Issues are organized in questions which can be defined 
by a committee or by means of a self-organized process internal to the community. The answers of each individual $j$ are organized in a vector $v^j$,  composed by $N_i$ cells. Each cell assume the value $1$ for a positive answer, $-1$ for a negative one 
or $0$ for a question left unanswered. Given the previous steps, the representative of  a given individual $j$ is selected by means of the following algorithm.

We compute the vector's overlap of each individual $j$ with all his neighbors $k$ through the following expression\cite{nos}:
\begin{equation}
v^j*v^k=\frac{\sum_{m=1}^{N_i} (v^j_m \cdot v^k_m)\delta(v^j_m, v^k_m)}{\sum_{m=1}^{N_i}(v^j_m \cdot v^k_m)^2}\,\, ,
\label{eq:FormerOverlap}
\end{equation}
where the numerator counts the number of questions answered in the same way (only yes or not) and the denominator counts the number of questions answered simultaneously by both individuals; $\delta$ stands for the Kronecker delta which is 1 if $v^j_m=v^k_m$
and 0 otherwise. Then, we calculate the
product of the previously defined overlap with the variable $I_{jk}$
(\textit{i.e.}, the integrity of $k$ as perceived by $j$), obtaining the ranking function:
\begin{equation}
R_{jk}=I_{jk}\, (v^j*v^k) \,\, .%=I_{jk}\, \frac{\sum_{m=1}^{N_i} v^j_m \, v^k_m\, \delta(v^j_m, v^k_m)}{\sum_{m=1}^{N_i}(v^j_m \, v^k_m)^2}\,\, ,
\label{eq:overlap_Demo}
\end{equation}

The introduction of the term $I_{jk}$ establishes a form of direct choice of the individual $j$ over the possible elected representative. In fact, overlaps, which are not controlled by voters, are weighted by a term subjectively assigned by the individuals. Note that we are simply considering the term $I_{jk}$ and not any possible statistical measure of the different  $I_{jk}$ associated to each agent $k$. In this way, we are clearly loosing information but the main goal of this study is to describe the effects of a subjective
term on the voting rule and not to obtain a more efficient voting rule in detecting the best representatives. Finally, each individual $j$ will indicate as her representative the individual $k'$ for which $R_{jk'}$ is maximum. In the case where the same 
maximum value is shared by more than one individual, the one with a higher connectivity is selected as the representative. For the exceptional case of equal connectivity, the representative is randomly chosen between the equivalent ones.

\begin{figure}[h]
\begin{center}
\includegraphics[width=0.8\columnwidth, angle=0]{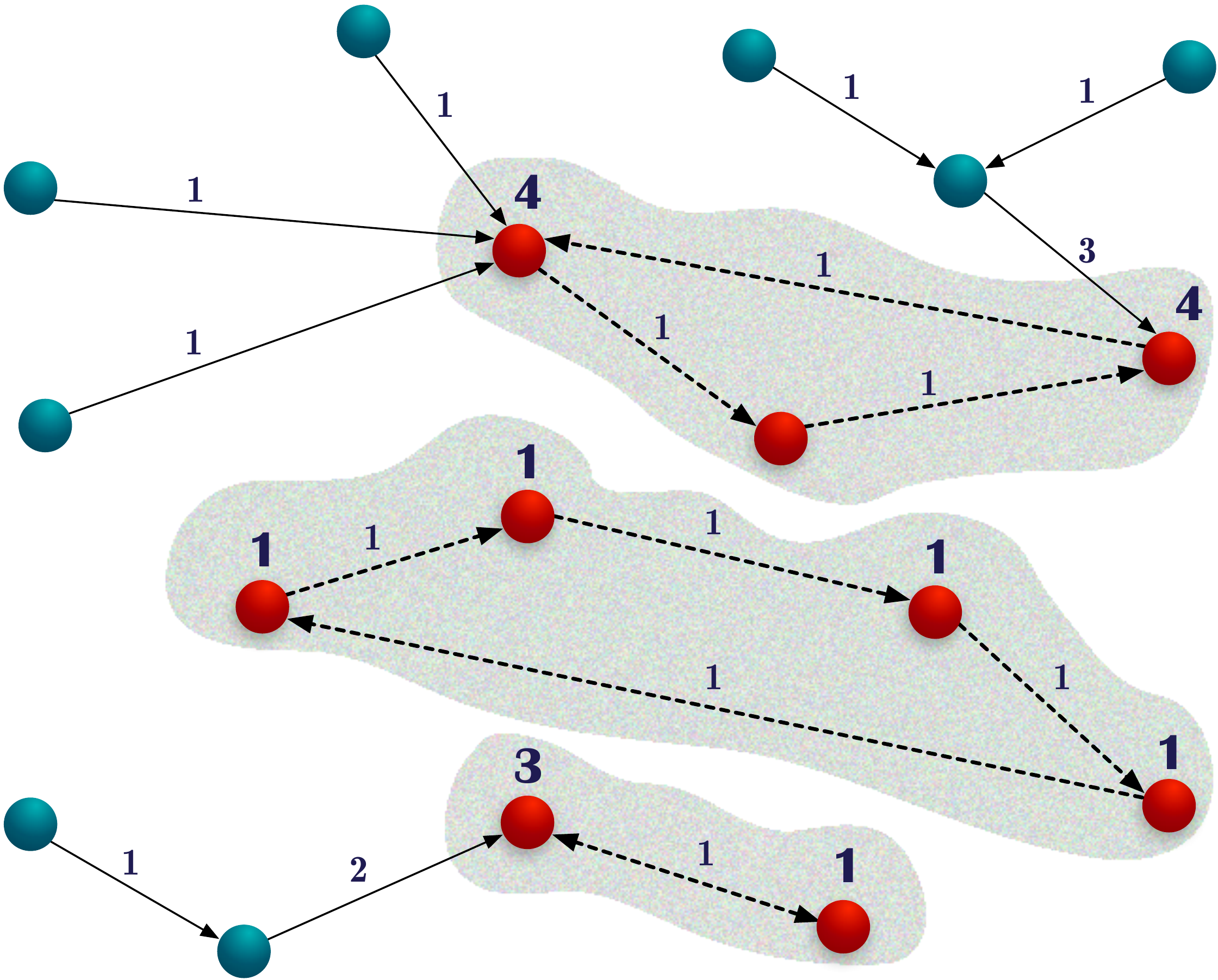}
\end{center}
\caption{Schematic representation of the vote process. 
Nodes stand for the individuals; the red ones belong to a cycle and will be confirmed as representatives if they collect more votes that the established threshold. The big numbers associated to the nodes represent the received cumulated votes. Arrows stand for the indication of each individual and the small numbers associated to them represent the number of transferred votes. Dotted arrows belong to a cycle, where there is no cumulative transfer of votes.}
\label{fig_Vote_Flux}
\end{figure}

After the selection of the representative $k'$ for every voter $j$, the final step consists of choosing the aggregate of representatives of 
the entire community. To this end, we  construct a directed graph, which we call the representative graph, where a node represents each individual and a directed link connects the individual with her personal representative. In this graph, which in general is composed by different disconnected clusters, cycles are present. These cycles represent individuals that have been mutually indicated by themselves. In details, we can affirm that the representative graph is a directed graph with out-degree 1. It is made of some disconnected components each one formed by a cycle with trees attached to the cycle nodes (see Figure \ref{fig_Vote_Flux}). As all the individuals outside the cycles are represented by the individuals belonging to them, individuals who belong to cycles are the proper potential representatives for the community. 

As a final step, among the individuals belonging to a cycle, only the ones with a number of votes larger than a threshold $\Theta$ are indicated as representatives. Votes are counted considering the cumulative flow defined by the directed graph. If the individual $j$ is pointing to $z$, $z$ receives all the votes previously received by $j$ plus one. This flow of votes is computed only following links outside the cycles. Inside the cycles, only the single vote of an individual is counted. To sum up, the votes $v$ received by an individual $i$ inside a cycle are equal to:
\begin{equation}
v_i=1+\sum_{t\in G(i)} l_t 
\end{equation}
where $G(i)$ is the set of all the trees ending at node $i$ and $l_t$ is the number of links of the tree $t$. Based on this score, the number of representatives is reduced and results to be a fraction of the total number of individuals that belongs to the cycles.

\section{Results}

In our simulations, each individual is assigned an intrinsic integrity 
$i_k$ which is a number  uniformly distributed in the interval $[0,1]$.
The perceived integrity $I_{jk}$ corresponds to $i_k$ shifted by 
the error in the perception that individual $j$ have on the integrity of individual $k$,
which is modeled by a scalar $\delta i_{j,k}$ drawn from a Gaussian
distribution $N(0,\sigma_p)$. In order to keep $I_{jk} \in [0,1]$,
$I_{jk}$ values greater than $1$ are set to 1 and negative values are set to $0$:
$I_{jk}=max[min(i_k+\delta i_{j,k},1),0]$. On the other hand, the individuals' opinions in relation to the selected issues 
are randomly generated with the following rule: given an issue $i$,
an individual does not have an opinion ($v_i= 0$) with probability $1/3$. 
The probability to have an opinion $v_i=+1(-1)$, is $1/3+\epsilon_i$ $(1/3-\epsilon_i)$, 
where $\epsilon_i$ is a random variable following 
a normal distribution with mean value equal to zero 
and $\sigma^2=0.05$. 

\begin{figure}[!t]
\includegraphics[width=0.49\columnwidth, angle=0]{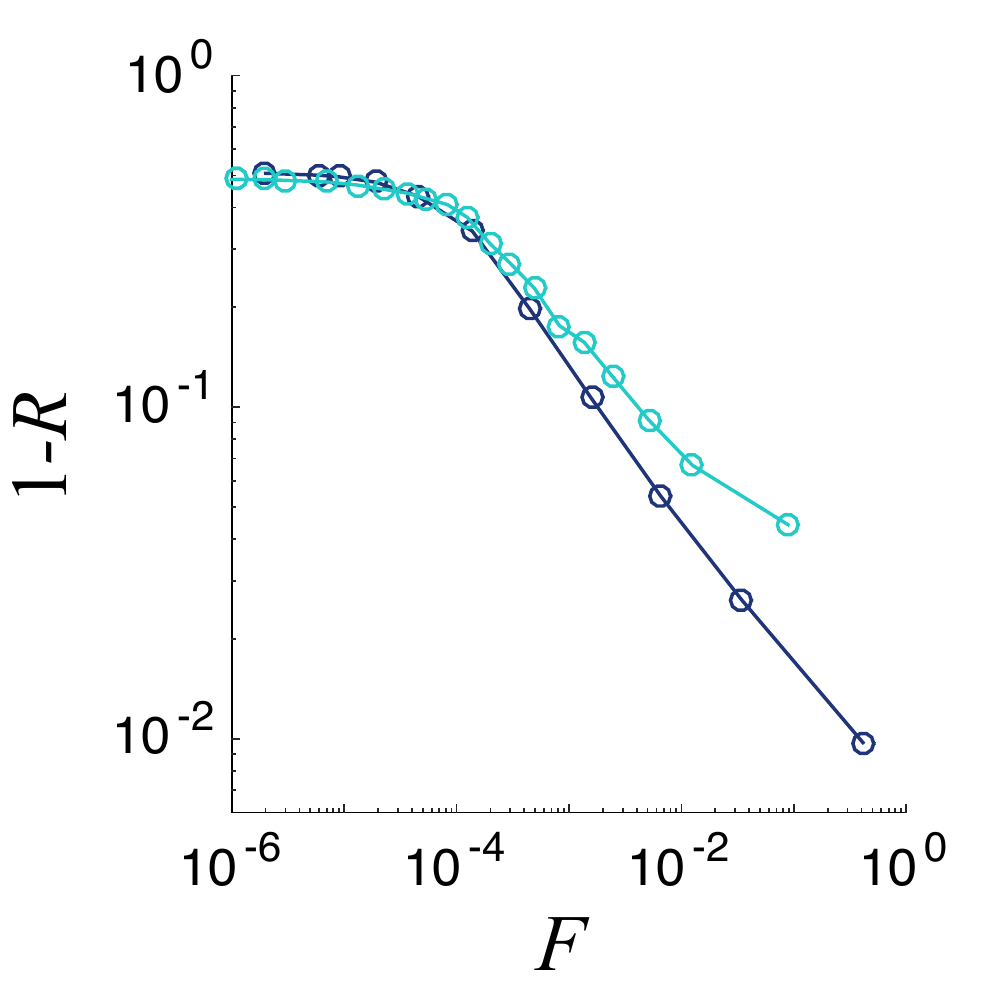}
\includegraphics[width=0.49\columnwidth, angle=0]{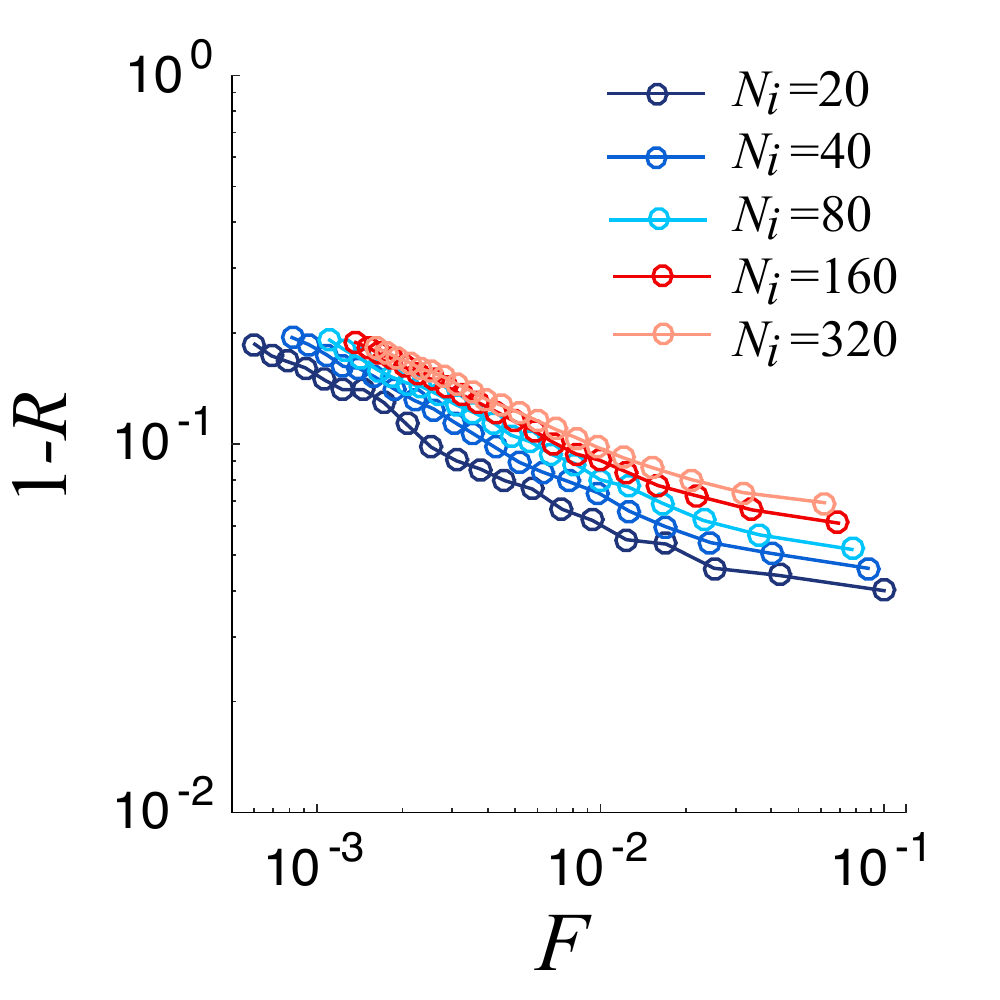}
\includegraphics[width=0.49\columnwidth, angle=0]{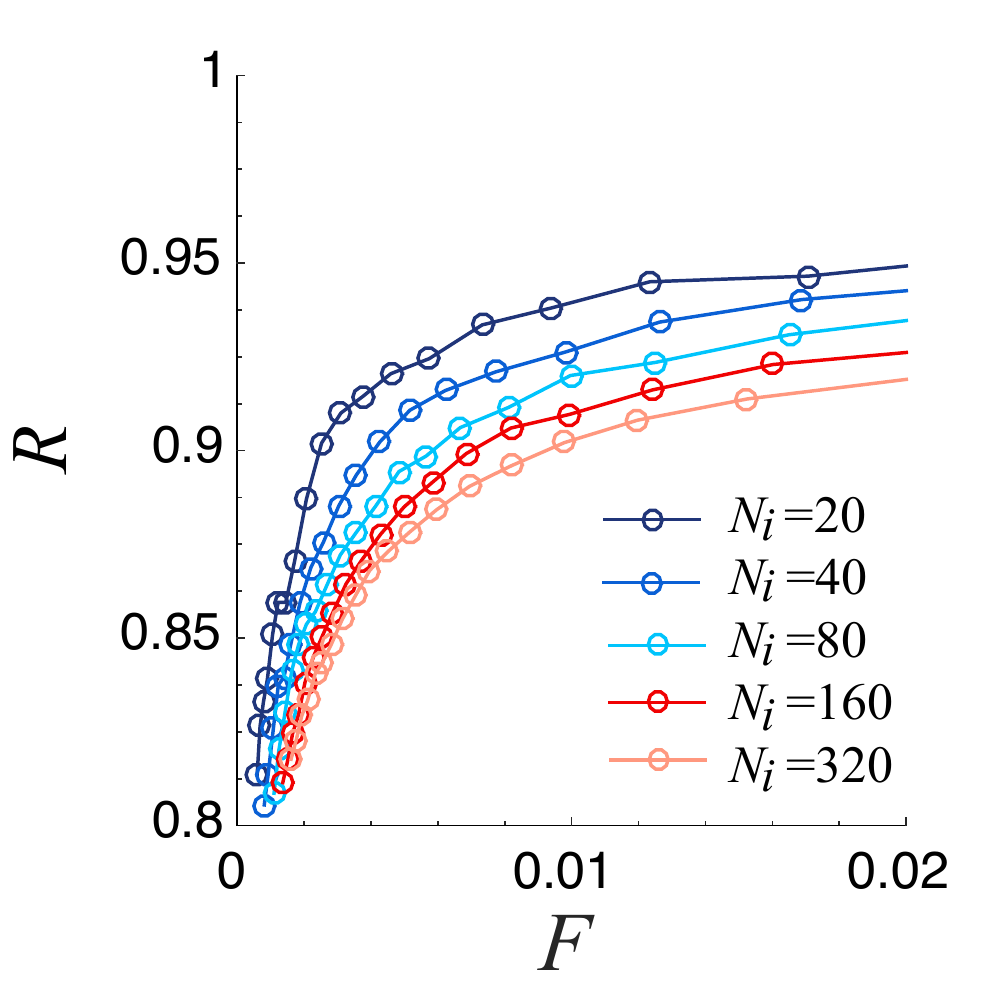}
\includegraphics[width=0.49\columnwidth, angle=0]{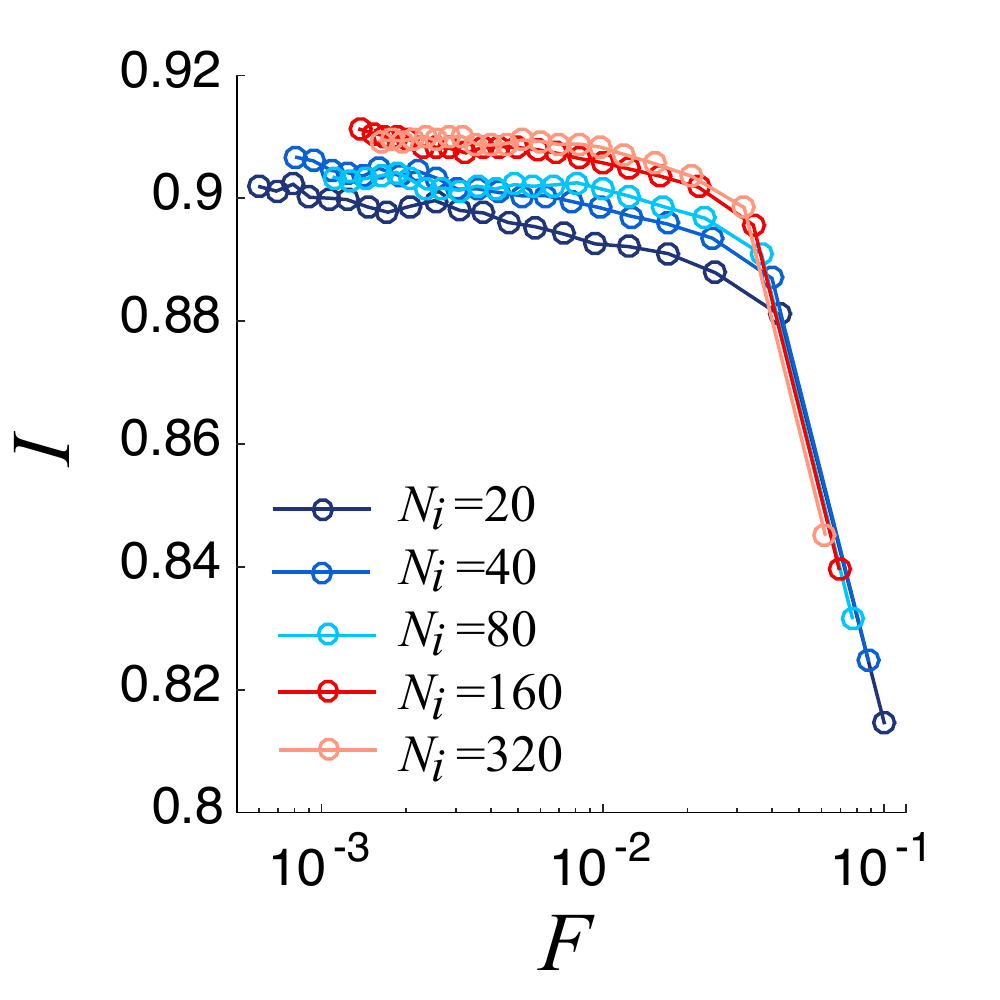}
\caption{{\it Top:} On the left, logarithmic plot of $1-R$ versus normalized committee size for the $NVR$ proposed in \cite{nos} (dark blue points) and the one proposed here ($NVR_I$, light blue points) 
with $Ni=40$. On the right,
 $1-R$ versus normalized committee size  for different $N_i$ values. {\it Bottom:}  Representativity (left) and Mean Commitee Integrity (right) as a function of normalized committee size. We consider a Erd\"os-R\'enyi network with $N_e=10000$, $\langle k \rangle =40$ and $\sigma^2=\sigma_p^2=0.05$ .Results are averaged over $100$ different realizations.}
\label{fig:RforNi}
\end{figure}

\begin{figure}[!t]
\includegraphics[width=0.49\columnwidth, angle=0]{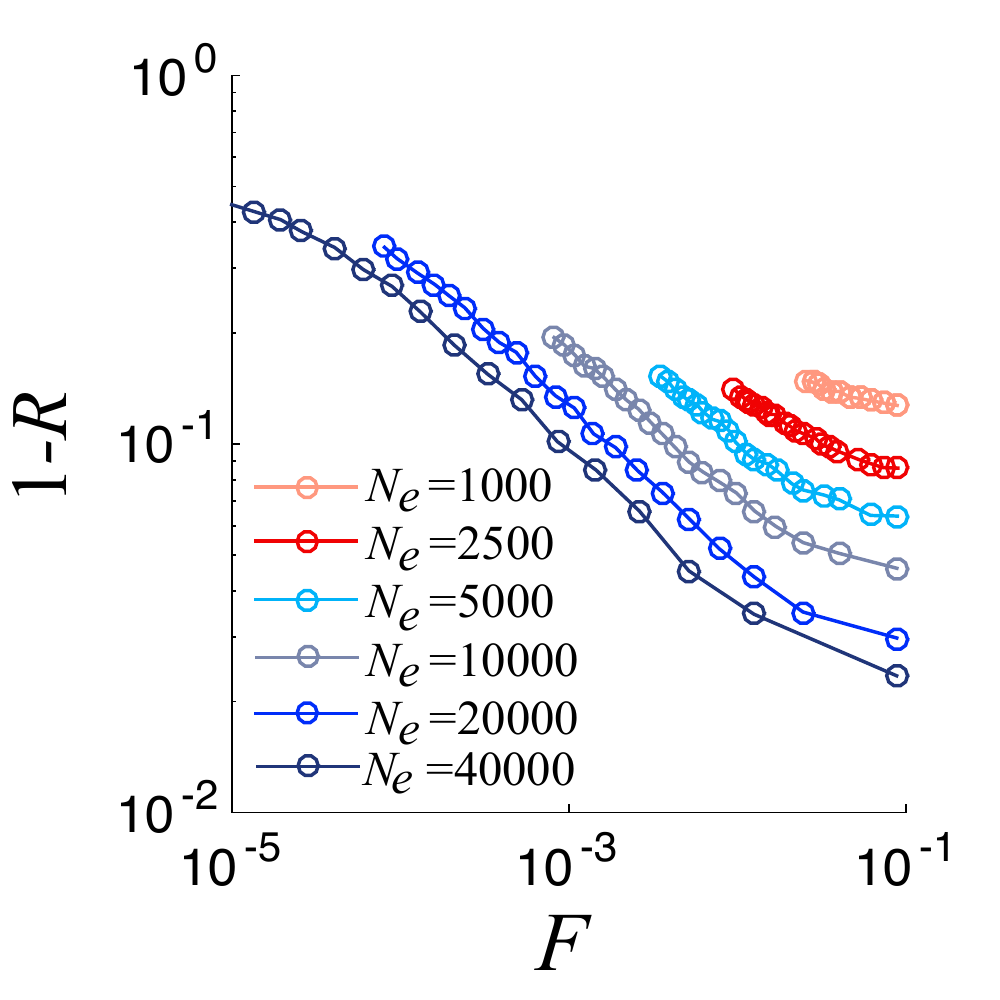}
\includegraphics[width=0.49\columnwidth, angle=0]{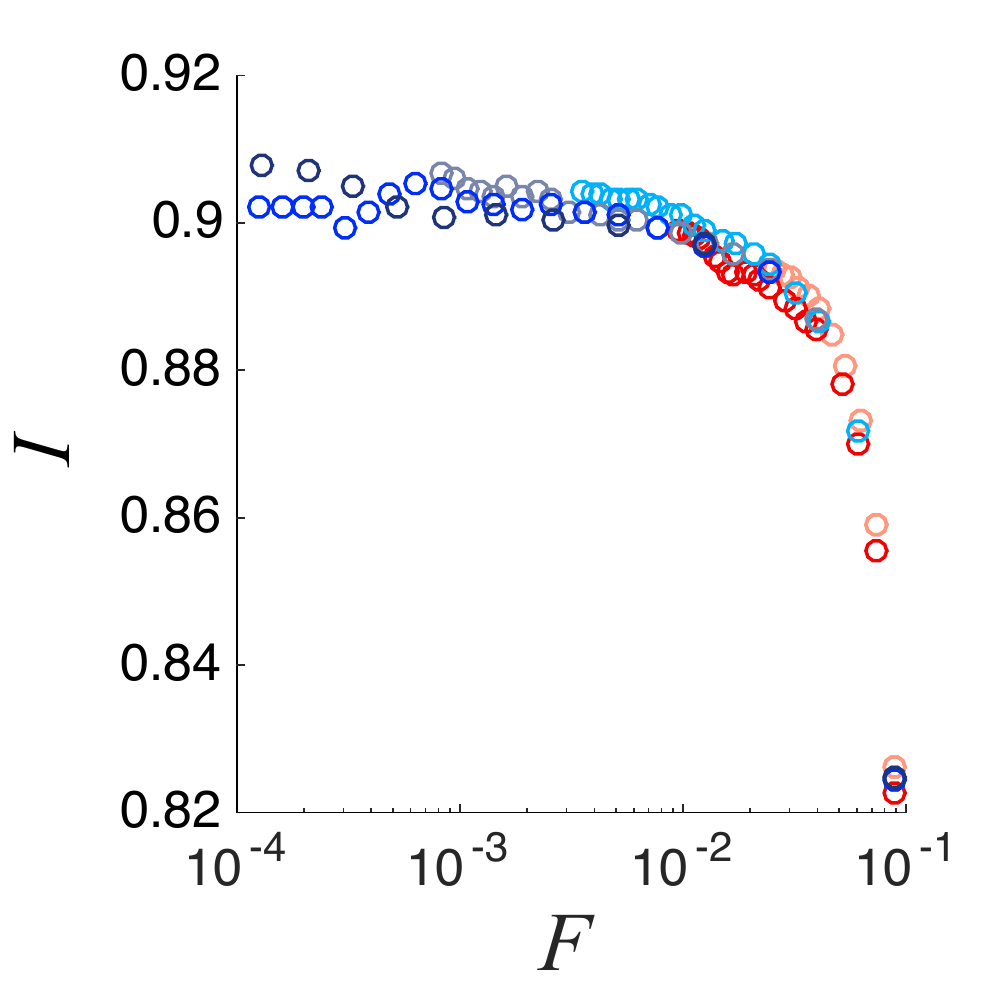} 
\caption{Logarithmic plot of $1-R$ versus normalized committee size (left). Mean Commitee integrity as a function of normalized committee size (right), for different numbers of electors $N_e$. We consider a Erd\"os-R\'enyi network with $N_e=10000$, $N_i =40$ and $\sigma^2=\sigma_p^2=0.05$. Results are averaged over $100$ different realizations.}
\label{fig:RforNe}
\end{figure}

The interaction circle of each individual is modeled 
generating a network where nodes represent individuals and links 
the social relationships present in the community.
The interaction circle of an individual is obtained
selecting a node and considering its first neighbors.
Note that an important simplification of this approach 
is the fact that it generates individuals with
symmetric social relationships. 
In the following analysis three types of networks are considered.
Homogeneous random networks, 
implementing the Erd\"os-R\'enyi model \cite{erdos}, where
the degree distribution is peaked around a typical value $\langle k\rangle$; heterogeneous networks,
using the Barabasi-Albert model \cite{barabasi},
with a power-law degree distribution $P(k) \propto k^{-3}$; and the so-called small-world Watts and Strogatz network model \cite{small}.
Our aim is not to model  specific aspects
of a real social network, but to use
simple examples just to discuss the possible 
influence  of some 
relevant network properties (such as the heterogeneity in 
the degree distribution, the average degree and the small-world property), on the behavior of our model.

The system can be characterized by three observables:
\begin{itemize}
\item The normalized committee size,  which is the 
ratio between the number of elected individuals ($E$) and 
the total number of individuals of the community:
$F=E/N_e$. 
%In order to have manageable committees it is preferable a small value of $F=E/N_e$.
%%%%%%%%%%%%%
\item The representativeness $R$, which is measured
calculating the fraction 
of decisions expressed by the elected 
committee ($e_j$) which matches with the community
decisions ($c_j$) over all the considered $N_i$ issues:
$R=\frac{\sum_{j=1}^{N_i} \delta(e_j-c_j)}{N_i}$.
The decisions of the elected 
committee are attained  by means of a majority
vote where each representative's vote is weighted 
by the numbers of popular votes he collected
during the election procedure.
The community decision correspond to the result of a plebiscite, % a direct process (plebiscite), 
where every individual votes following the opinion
expressed in his vector $v^j$ (no opinion corresponds to
abstention). %Note that if the individual has no opinion on a particular issue, he abstains from voting.
For $R=1$ a perfect representativeness is obtained: a committee 
makes all the decisions in line with the popular will.  
On the opposite, for binary decisions,  $R=1/2$ corresponds to a non-representative committee, whose decisions are completely uncorrelated to the popular will.
A useful observable is $1-R$, which measures how far the system
is from the perfect representativeness.
This quantity is particularly interesting because, for the original
model without integrity \cite{nos},
it presents a simple and robust relation with $F$:
\begin{equation}
1-R \propto 1/\sqrt{F}
\label{result}
\end{equation}
%%%%%%%%%%%%
\item The integrity $I$ which is the mean value of the 
intrinsic integrity $i_k$ of the individuals selected for the committee.\\
\end{itemize}

We perform our analysis varying the value of the threshold $\Theta$, such as to obtain 
committees of relatively small size but
which express a high level of representativeness - close to 0.9 - (see \cite{nos} for details).
In order to explore the relation between  committee  size and 
representativeness we plot the representativeness versus 
the normalized committee size.
As can be clearly appreciate in the logarithmic plot of $1-R$ versus the normalized committee size, $F$ (Figure \ref{fig:RforNi}), the introduction of the integrity parameter has a marginal impact
on the relation \ref{result}, which is conserved also after the introduction of the selection of the individuals' integrity.
Only for higher values  of $F$, which are unpractical, a slightly worse representativeness in relation to the classical algorithm can be perceived.
 As for the classical algorithm, for fixed R, the normalized committee size increases if the number of issues is increased.
The integrity behavior as a function of $F$ has a quite simple response:
it shows very high values and a final abrupt drop for large committee size, because, in this situation, the probability for lower integrity individuals to obtain the amount of votes needed to be elected become relevant.
The dependence on $N_i$ is weak and establishes a tradeoff between Representativity and Integrity. More issues make the overlap less relevant in the computation of $R_{ik}$ improving the integrity at the expenses of the representativity.

\begin{figure}[!t]
\includegraphics[width=0.98\columnwidth, angle=0]{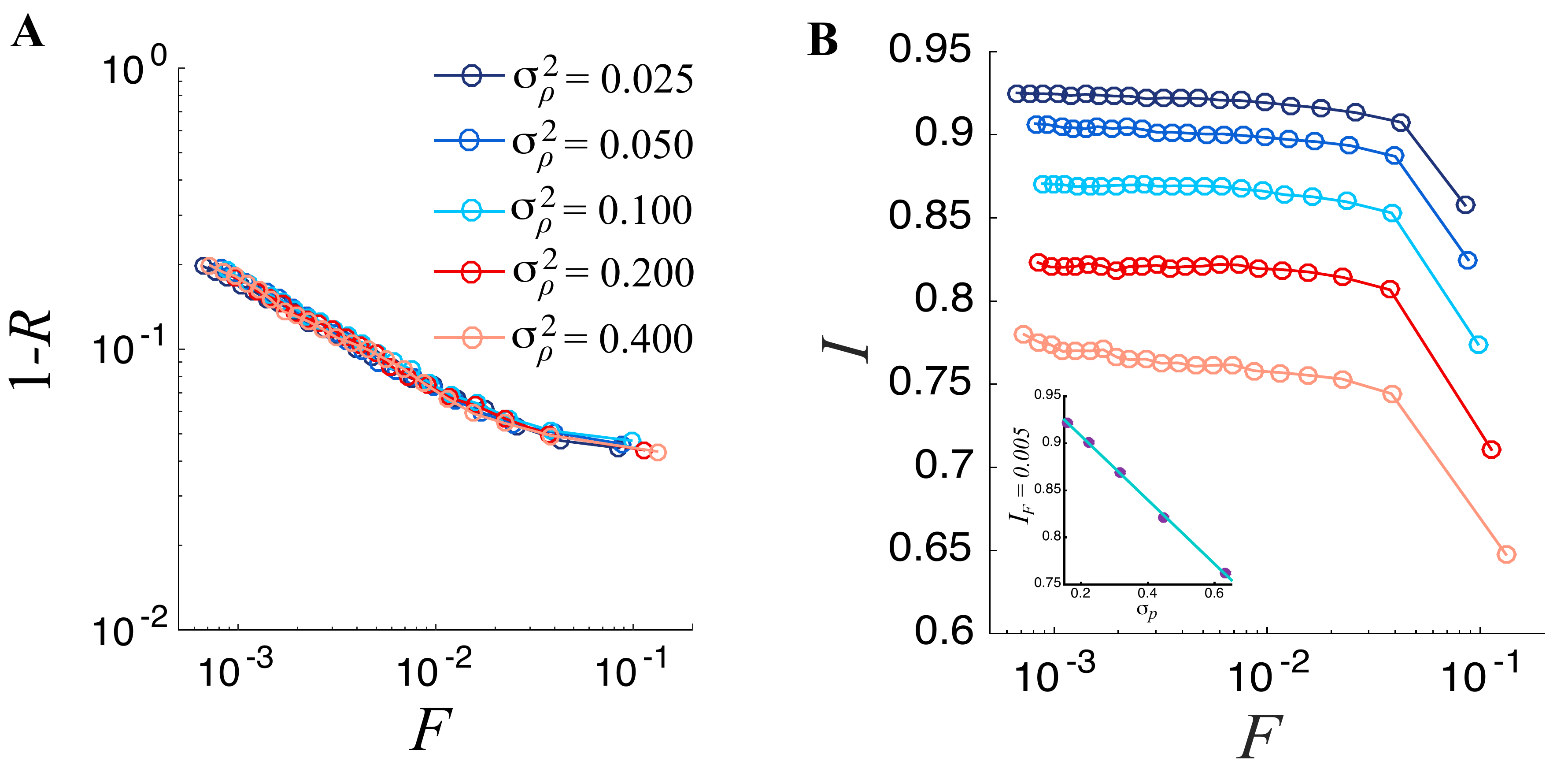}
\caption{
Logarithmic plot of $1-R$ versus normalized committee size (left). Mean Commitee Integrity as a function of normalized committee size (right). In the inset we show the linear behaviour of $I_{F=0.005}$ with $\sigma_p$ ($I_{F=0.005}=-0.34 \sigma_p+0.98$). We consider a Erd\"os-R\'enyi network with $N_e=10000$, $N_i=40$, $\sigma^2=0.05$ and $\langle k \rangle =40$. Results are averaged over $100$ different realizations.}
\label{fig:RforSigma_p}
\end{figure}

The dependence of the above observables with the system size $N_e$ (Figure \ref{fig:RforNe})
shows that the latter has an impact on the representativeness but not on the integrity behavior.
In fact, as it was the case in the original model, when fixing R the committee size decreases for larger 
system sizes. For example, for the parameters used in Figure \ref{fig:RforNe}, a representativity of 0.9 corresponds to a committee of 78 members for a community of 2500 individuals, and to 36 representatives for $N_e=40 000$. Furthermore, as can be seen in Figure \ref{fig:RforSigma_p}, the error in the perception of the integrity, which is controlled by the parameter $\sigma_p$, has no effect on the  
representativeness. In contrast,  it obviously affects the committees' integrity. The plateaux values of $I$ decrease with $\sigma_p$, following a simple linear dependence on this parameter. Higher values of errors in the integrity perception linearly correspond to
worse values in the integrity selection (see inset in Figure \ref{fig:RforSigma_p}).

\begin{figure}[!t]
\includegraphics[width=0.49\columnwidth, angle=0]{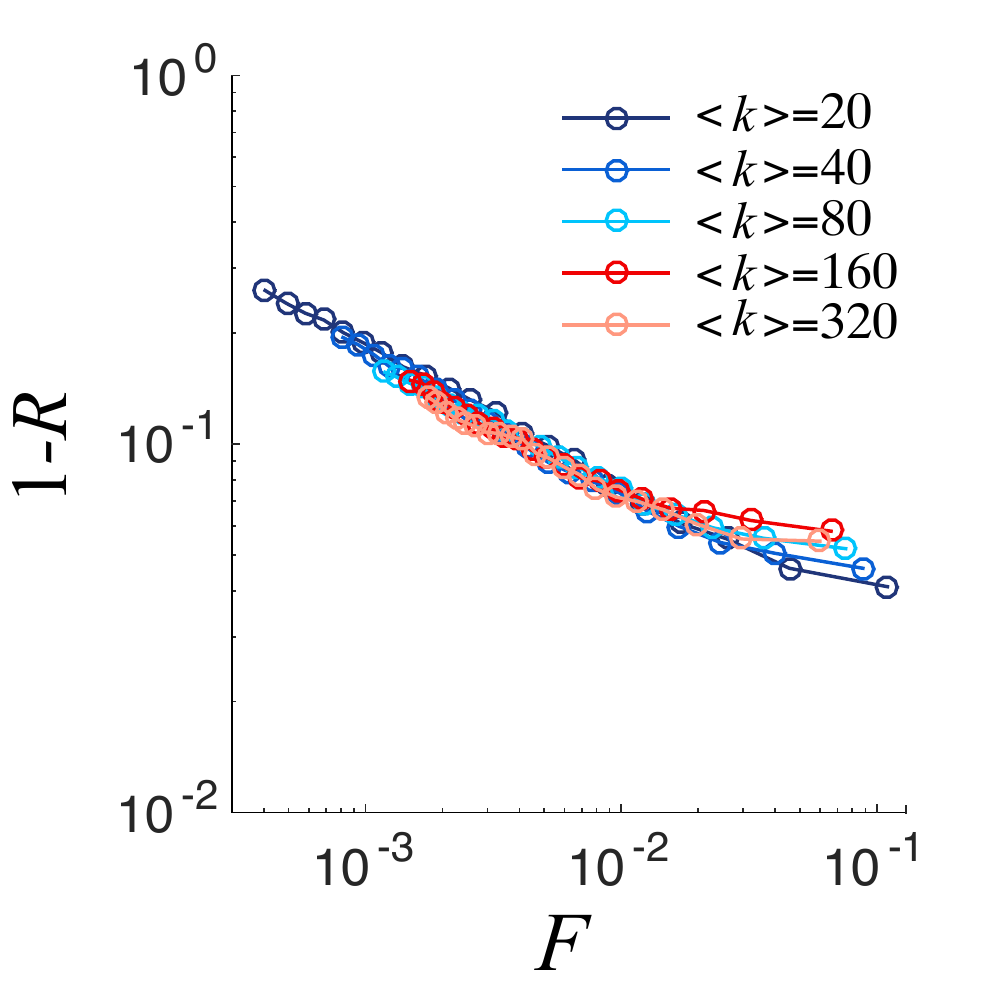}
\includegraphics[width=0.49\columnwidth, angle=0]{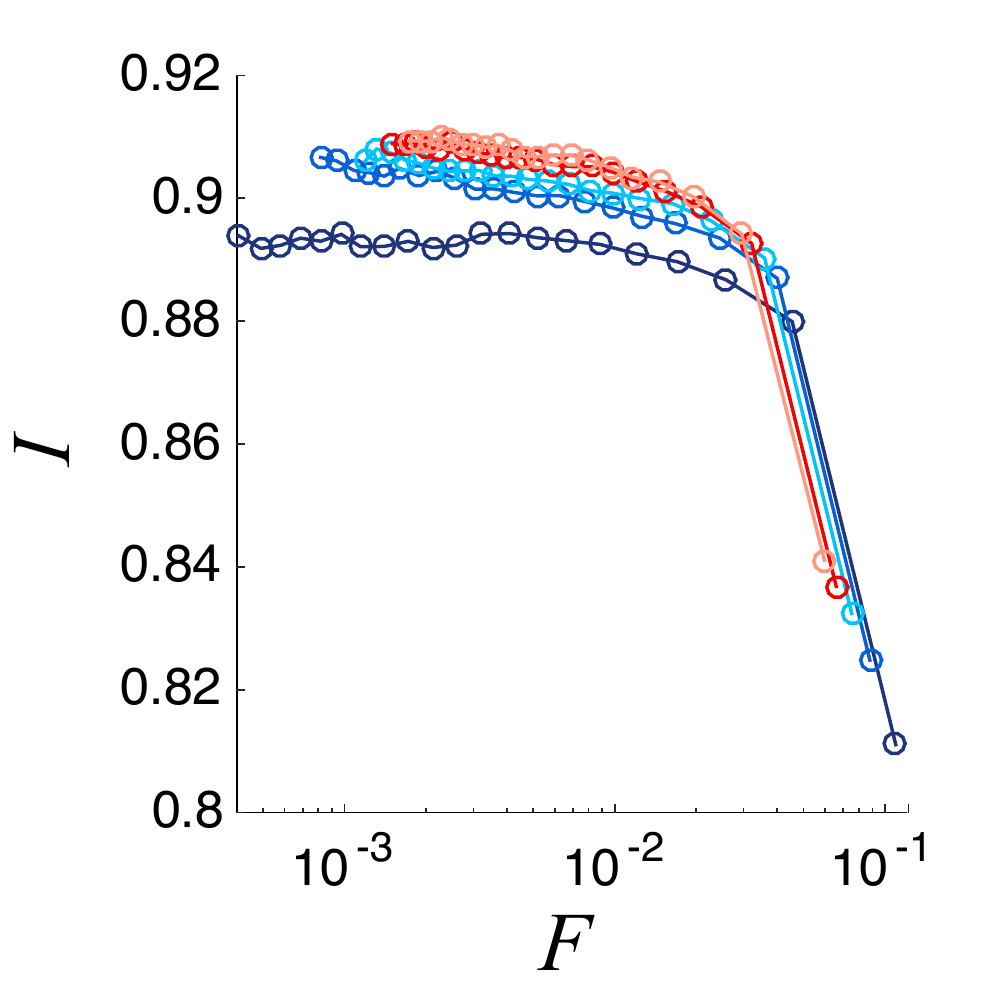} 
\includegraphics[width=0.49\columnwidth, angle=0]{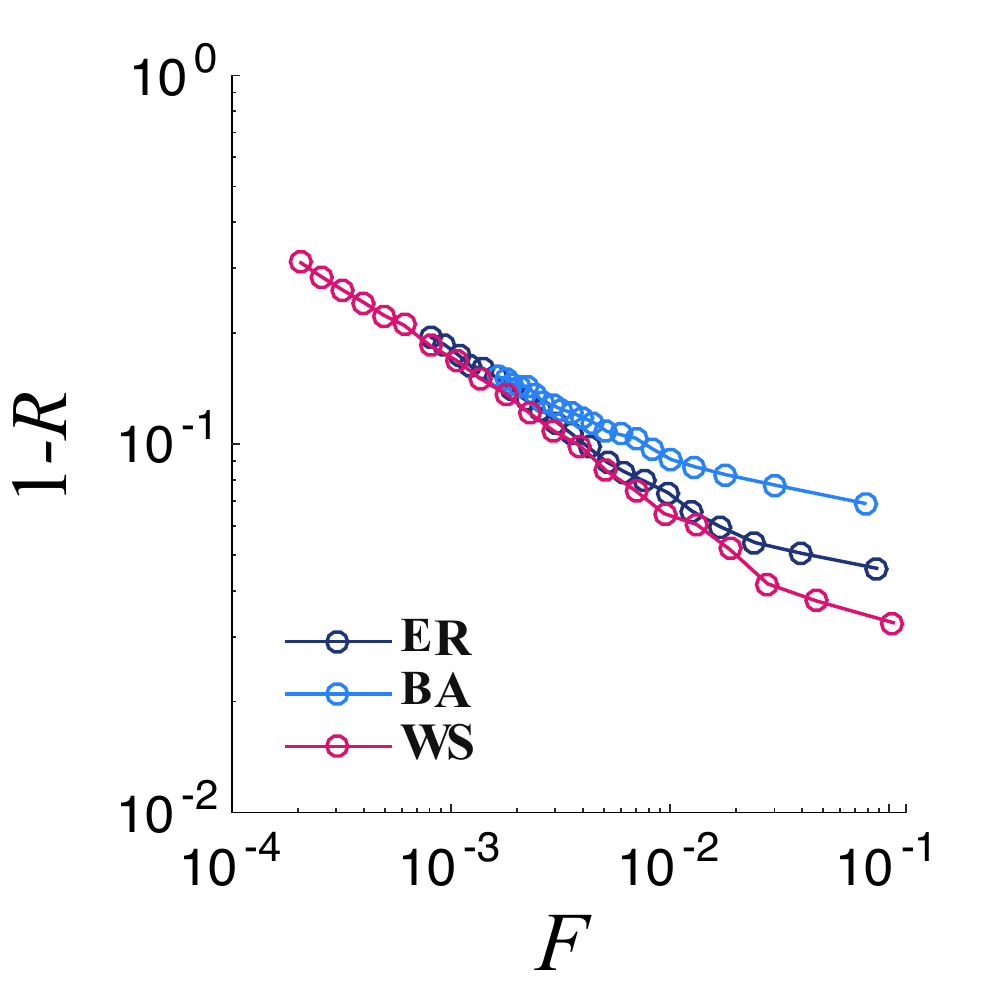}
\includegraphics[width=0.49\columnwidth, angle=0]{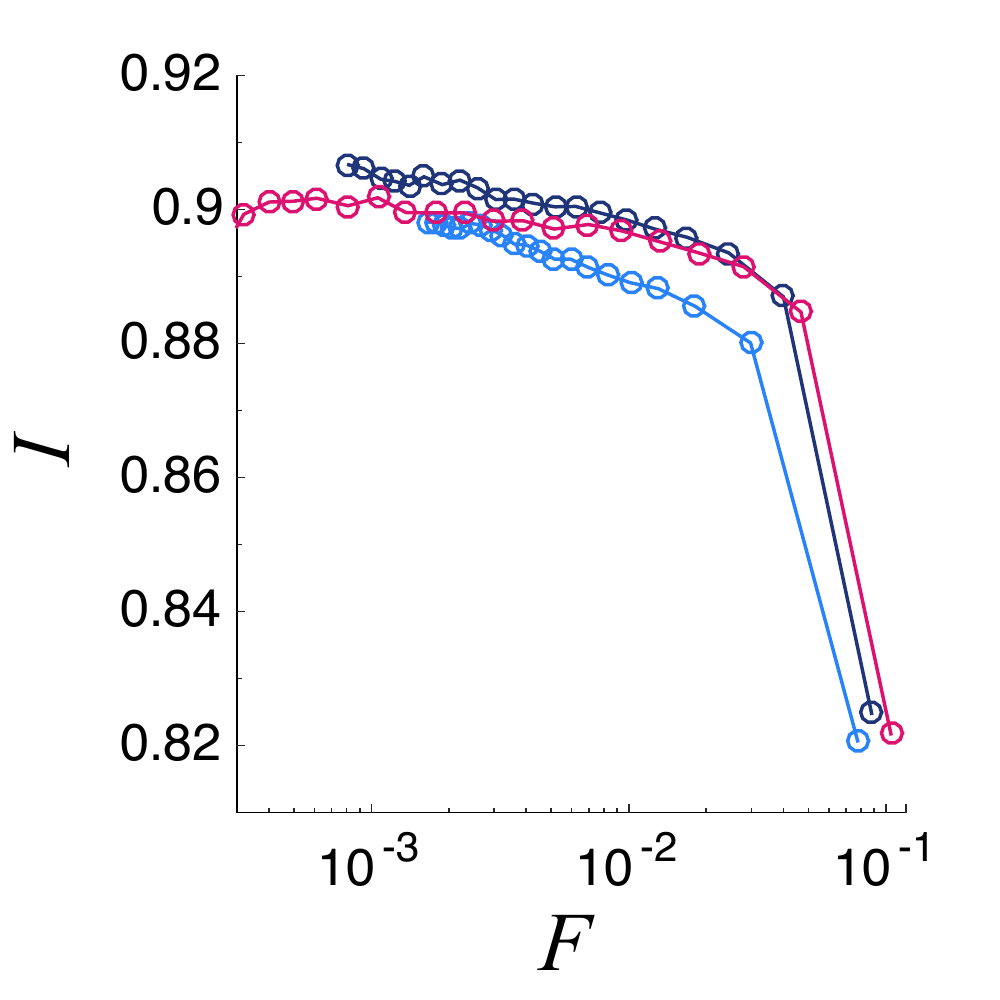}
\caption{Top: Logarithmic plot of $1-R$ versus normalized committee size (left) and Mean Commitee Integrity as a function of normalized committee size  (right) for different values of $\langle k \rangle$. Bottom: Logarithmic plot of $1-R$ versus normalized committee size (left) and Mean Commitee Integrity as a function of normalized committee size  (right) for different network topologies. We consider a Erd\"os-R\'enyi network with $N_i=40$, $\langle k \rangle =40$ and $\sigma^2=\sigma_p^2=0.05$. Results are averaged over $100$ different realizations.} 
\label{fig:RforNetw}
\end{figure}

In Figure \ref{fig:RforNetw}, we can see that the representativeness is not strongly dependent on the connectivity of the network. For sufficient high $\langle k \rangle$, the curves show the same behaviors. The heterogeneity in the degree distribution of the network  marginally impacts the results. For high values of $F$, the Barabasi-Albert network performs moderately worse than the Erd\"os-R\'enyi one. As in the case of the original algorithm, higher connectivity generates a small bias in the selection of the more representative individuals. In contrast, the small world property of the Watts-Strogatz network positively influences 
the algorithm allowing slightly better results in terms of representativeness. This last behavior is more pronounced than in the case of the original algorithm.

\begin{figure}[!t]
\includegraphics[width=0.49\columnwidth, angle=0]{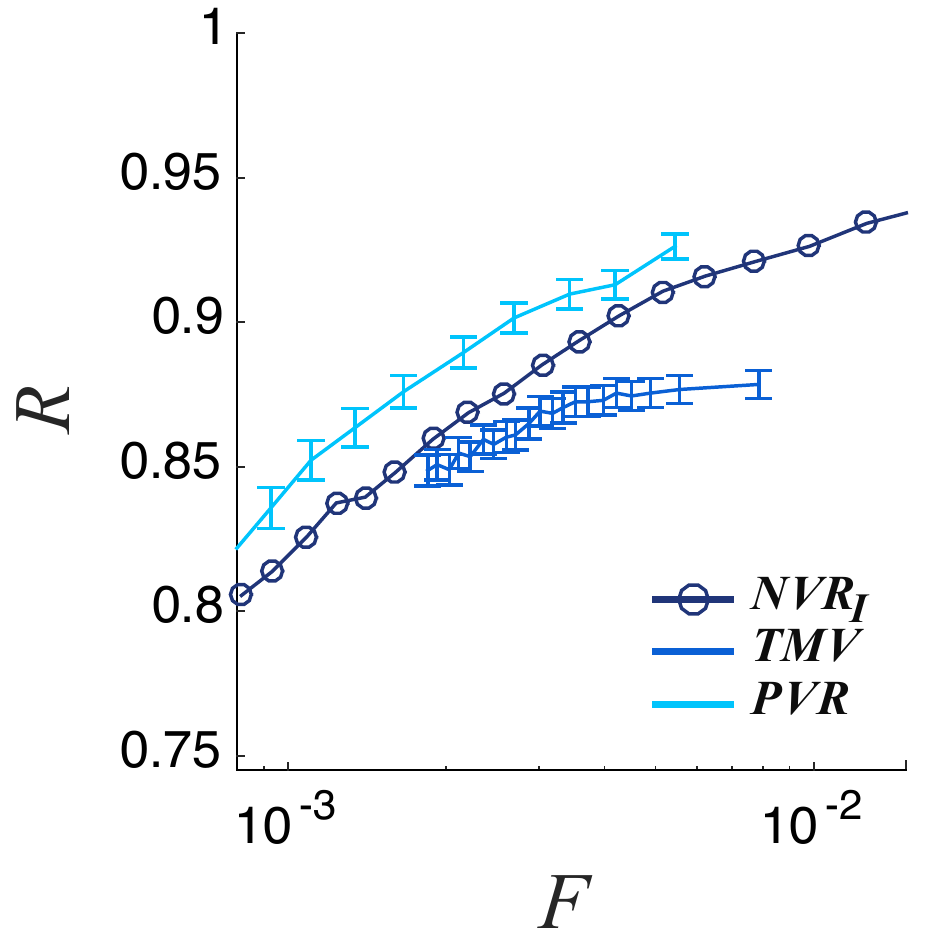} 
\includegraphics[width=0.49\columnwidth, angle=0]{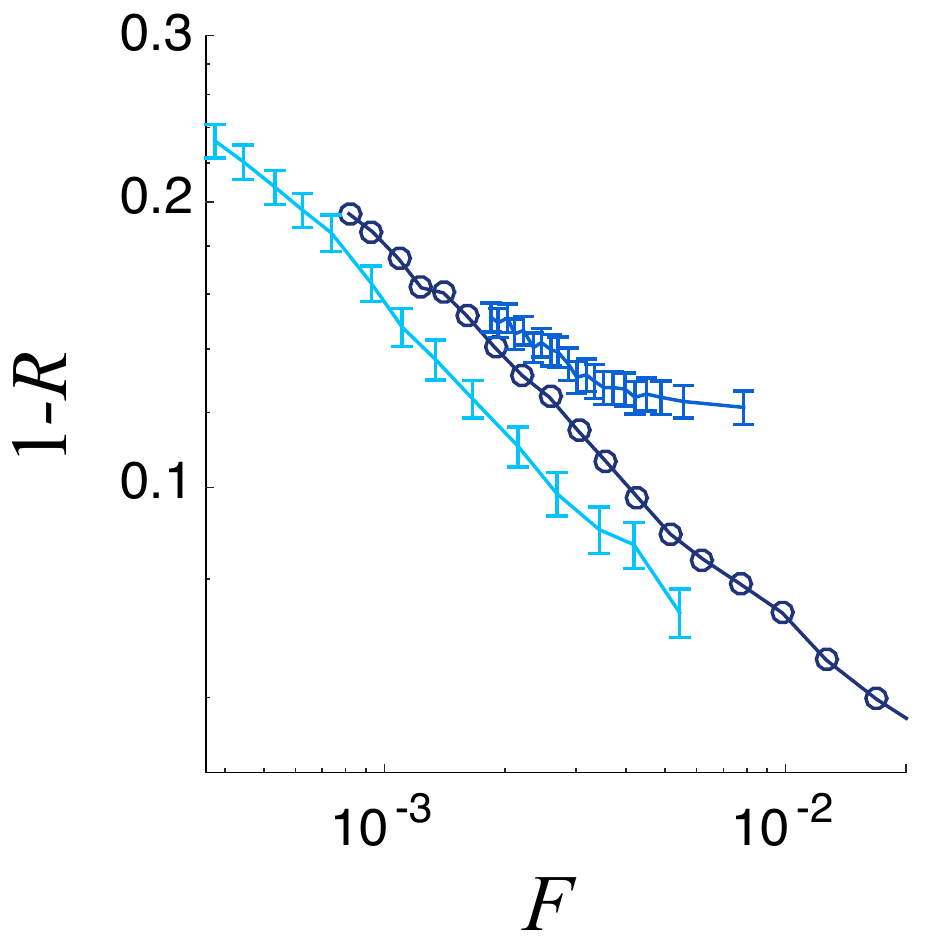} 
\includegraphics[width=0.49\columnwidth, angle=0]{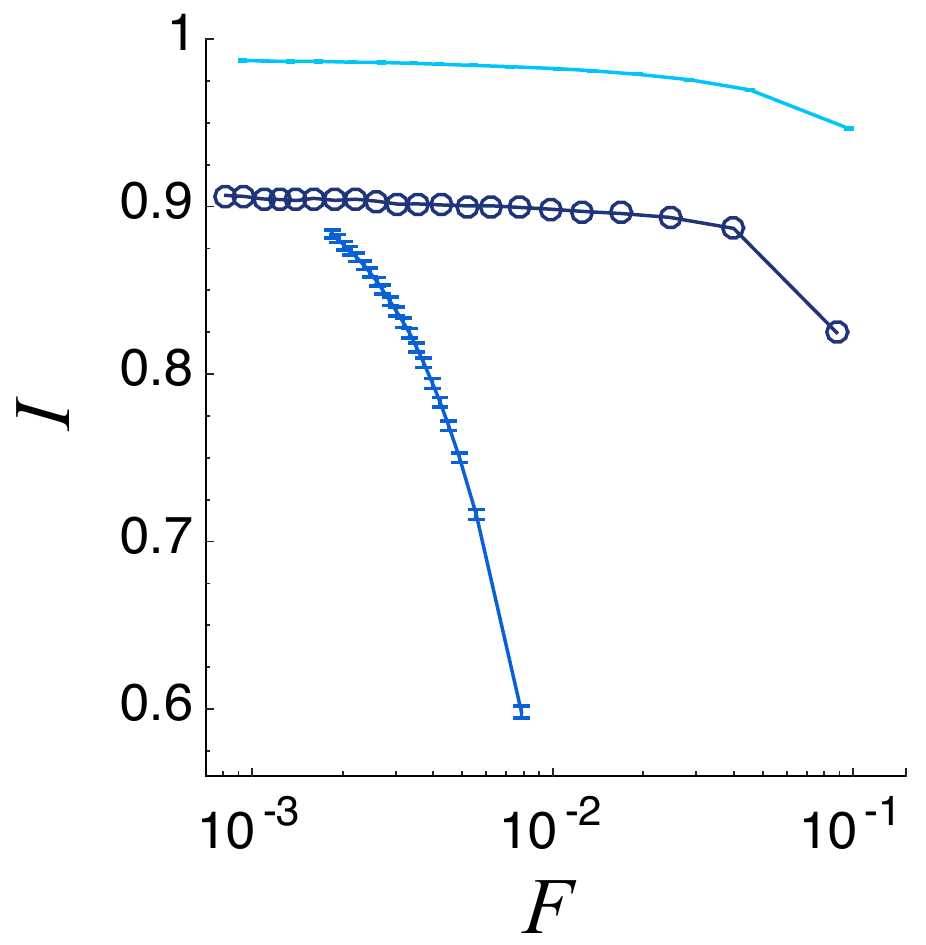}
\includegraphics[width=0.49\columnwidth, angle=0]{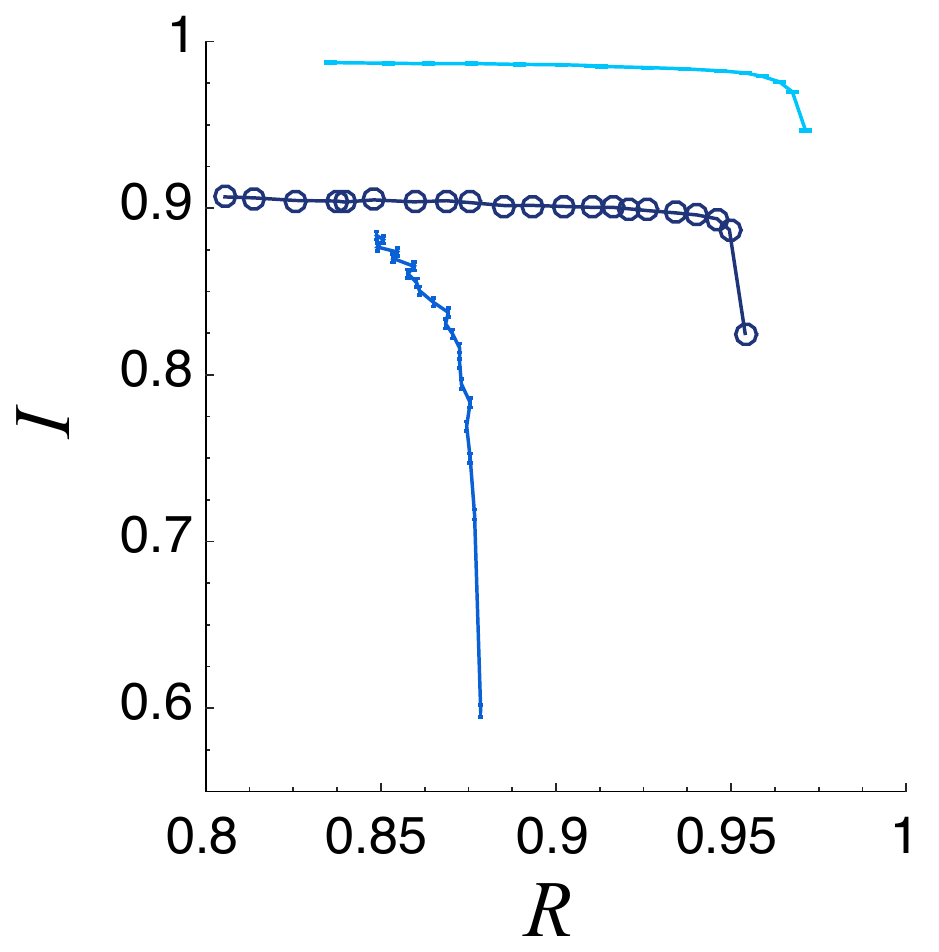}
\caption{Top: Representativity versus normalized committee size (left), logarithmic plot of $1-R$ versus normalized committee size (right). Bottom: Mean Commitee integrity as a function of normalized committee size (left) and Representativity (right). The results correspond to the Networked Voting Rule ($NVR_I$), the Traditional Majority Voting ($TMV$) and a Perfect Voting Rule ($PVR$) with parameters $N_e=10000$, $N_i=40$ and $\sigma^2 = \sigma_p^2 = 0.05$. For the $NVR_I$ we use a Erd\"os-R\'enyi network with $\langle k \rangle =40$. For the $TVR$ we set the initial number of candidates $N_c=100$. Results are averaged over $100$ different realizations. See the main text for a detailed explanations of the voting rules.}  
\label{fig:CompPoliSyst}
\end{figure}

An important analysis is the comparison of our model behavior
with other traditional methods of representatives selection. 
To this end, we compare theoretically the representativeness and 
the integrity of committees of the same size.
A first  widespread model is a traditional majority voting rule (TMV) for 
electing representatives in a closed list of previously determined candidates. 
In our implementation, a list of $Nc$ candidates is randomly selected among the community and each individual $j$ votes for the candidate who presents the higher $R_{jk^*}$ value 
($k^*$ belongs to the list of  $Nc$ candidates). Note that also in this case the evaluation of the integrity is influenced by errors in the perception. Decisions are taken with the same weighted voting rule. This modeling approach mimics a voter who has
a perfect
knowledge of the candidates, and it assumes that he makes a rational decision to maximize his representation. Also for this voting rule, representativeness is computed by comparing the decisions taken by the committee, obtained with a weighted majority voting process, with the results of the direct popular vote. As can be appreciated in Figure \ref{fig:CompPoliSyst}, our model is by far more efficient, reducing the size of the committees in more than a half and showing a better selection of the integrity of the representatives.

Finally, we compare our method to an idealized perfect voting rule (PVR). This rule represents a situation of rational individuals that have a perfect knowledge  of all the other individuals, which means that they perfectly know the opinion of all the other individuals. Moreover, they are globally networked, which means that they have a direct access to all other individuals, allowing their acts to be checked. In this situation, a voter indicates the individual that has the higher overlap with his opinion vector and the best integrity (the higher $R_{jk^*}$ value). Note that in this case the evaluation of the integrity is not influenced by errors in the perception.
The selected committee is formed by the first $F\cdot N_e$ individuals which poll more. 
Also in this case, the committee decisions are taken by means of a weighted majority vote. This voting rule, although unrealistic, is still useful, at least, in two respects. First, very small communities can exhibit similar characteristics. Second, the model is a useful yardstick for evaluating the levels of representativeness of other more realistic models. The relation between representativeness and committee sizes can be compared also in the case of the PVR  rule (see Figure \ref{fig:CompPoliSyst}). 
It is quite impressive that the representativeness of our networked voting rule is comparable with the perfect one. The PVR rule is able to select a committee with a higher integrity score, but this is possible because in this situation integrity of everybody is tested, and not 
only the integrity of a small subset, as it happens for our networked rule.

\begin{figure}[!t]
\includegraphics[width=0.49\columnwidth, angle=0]{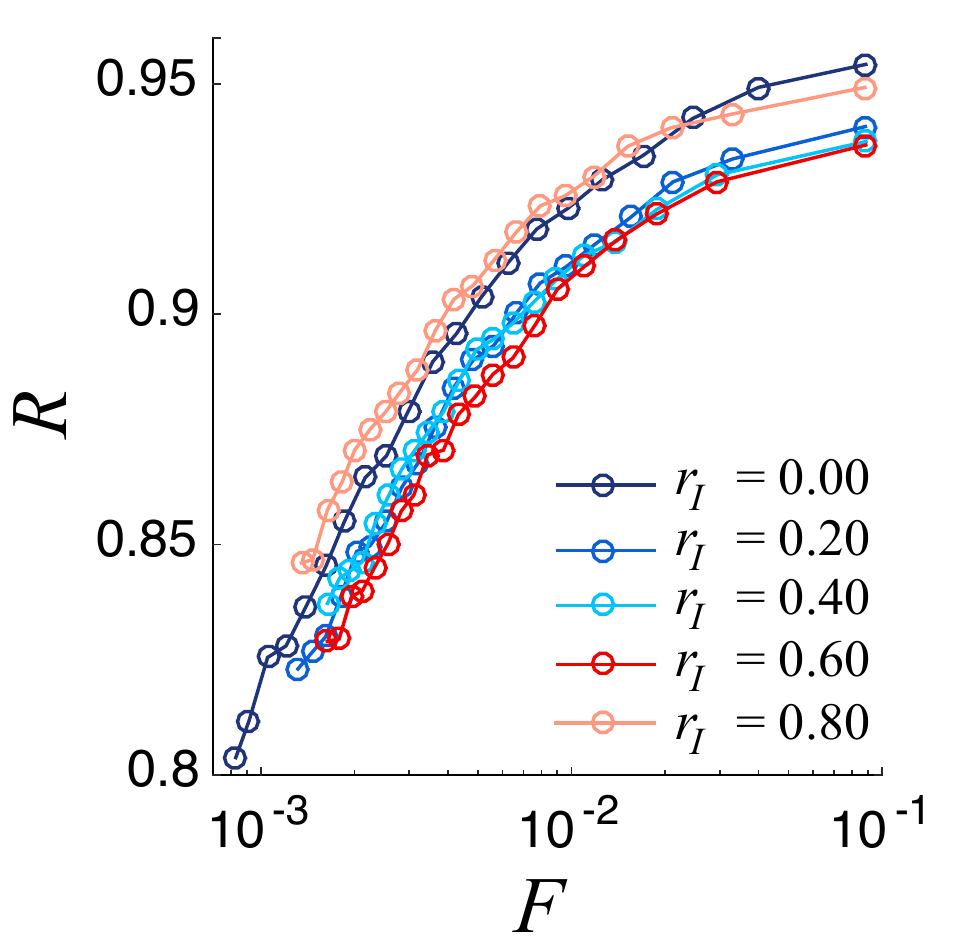} 
\includegraphics[width=0.49\columnwidth, angle=0]{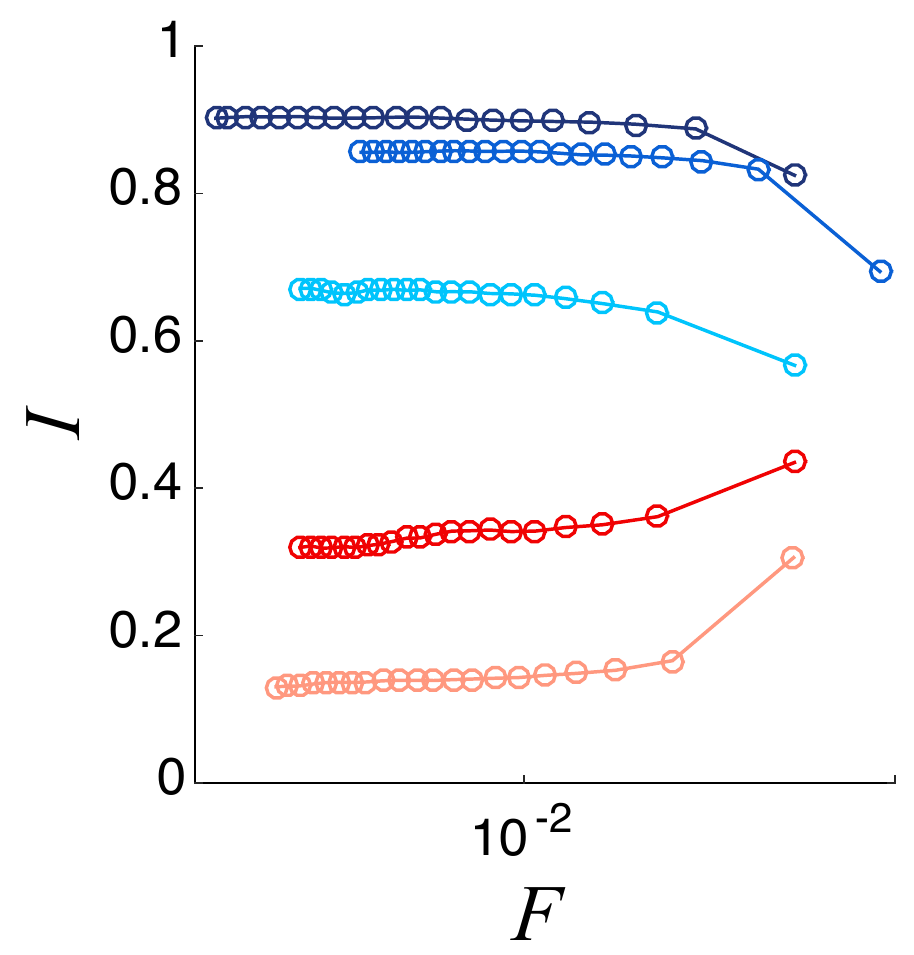} 
\caption{Top: Representativity versus normalized committee size (left), Mean Committee integrity as a function of normalized committee size (right). 
}
\label{fig:I_trans}
\end{figure}

\begin{figure}[!t]
\includegraphics[width=0.7\columnwidth, angle=0]{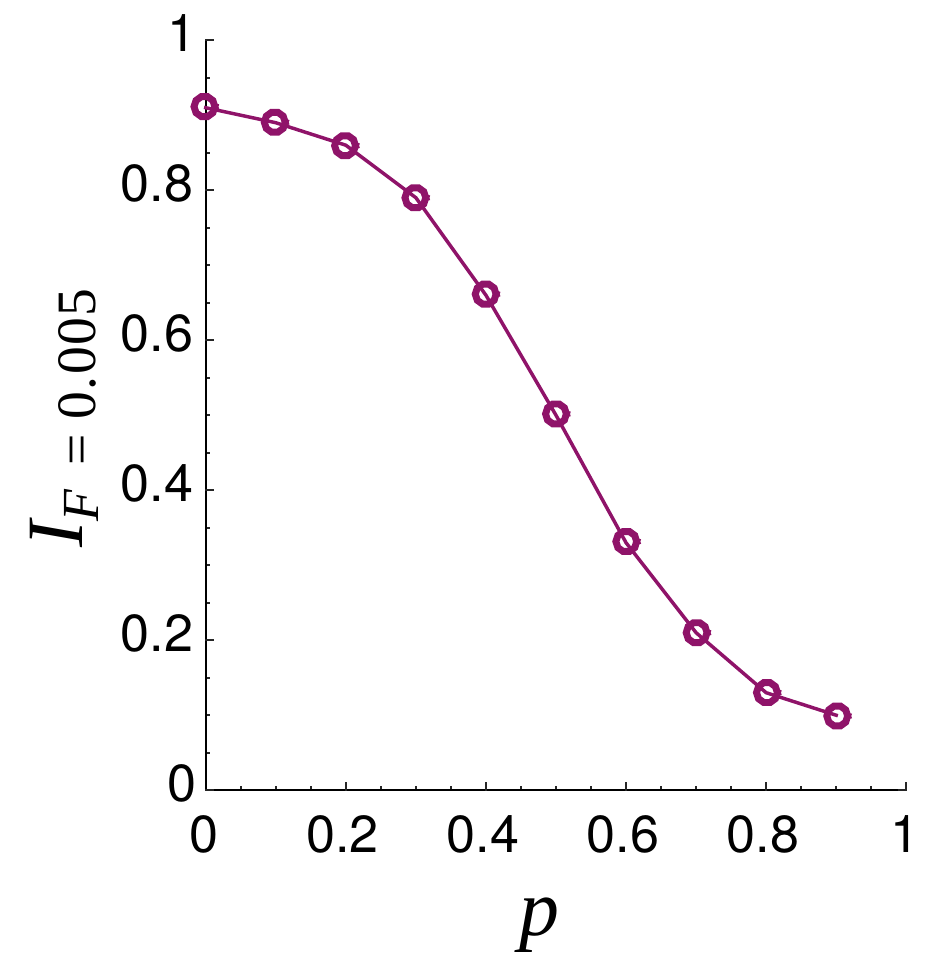}
\caption{Mean Commitee Integrity for a normalized committee size $F=0.005$ ($I_{F=0.005}$) as a function of the percentage of individuals with a distorted perception of the integrity ($p$). We consider a Erd\"os-R\'enyi network with $N_e=10000$, $\langle k \rangle =40$ and $\sigma^2=\sigma_p^2=0.05$. Results are averaged over $100$ different realizations.}
\label{fig:I_trans}
\end{figure}

We conclude our analysis testing the resilience of our
networked voting rule to possible attacks.
We consider the situation in which a group of voters decides
to assign high $I_{jk}$ scores to some individuals who,
in contrast, are characterized by a low personal integrity.
This behavior can model relations of patronage and clientelism, where individuals with low integrity organize a network of social
relationships for obtaining political support. 
In our model this behavior can be modeled introducing a percentage
of individuals $p$ for whom $I_{jk}=1-I_{jk}$.
As can be seen in Figure \ref{fig:I_trans}, representativeness
is not seriously affected by this actions. In contrast, the integrity
of the elected committee is obviously strongly influenced by this
ill behavior. As can be clearly seen fixing a value for $F$,
integrity undergoes an abrupt transition from high values
towards very low values as $p$ increases.

\section{Conclusions}

We analyzed a new voting algorithm,  particularly well suited for online social 
networks, for selecting a committee of 
representatives with the aim of enhancing  the participation of a community 
both as electors and as representatives.
This voting system is based on the idea of transferring votes 
through a path over the social network (proxy-voting systems). 
Votes are determined by an algorithm
which weights the similarity of individuals opinions and the trust between individuals directly connected in a specific social network.

Our computational analyses suggest that 
this voting algorithm can generate high representativeness for 
relatively small committees characterized by a high level 
of integrity. 
Results of representativeness and integrity are comparable with a theoretically defined perfect voting rule and, in general, they perform better than a traditional 
voting rule with a closed list of candidates.
The introduction of a term which expresses the trust on the integrity 
of a candidate does not significantly impact the representativeness of the 
committee, in particular for small and medium size committees.
The rule shows a robust behavior in relation to  the community size.
Besides, the perception of individual integrity directly influences the quality
of the committee: higher values of errors in the integrity perception linearly correspond to poorer values in the committees' integrity.
On the other hand,  representativeness is not strongly influenced by integrity perception.

Interestingly enough, these findings are not strongly dependent on the 
general properties of the network used to describe the community of voters,
as shown by the analysis of networks characterized by different topologies. Finally, the voting system seems robust to 
strategic and untruthful application  of the voting algorithm.
In fact, even with a $20\%$ of the votes produced by individuals which
vote for candidates with a low personal integrity,
 the integrity of the committee is  substantially unaltered and only if unfair votes
 are around $40\%$ of the votes an abrupt change is observed. In conclusion, we believe that the voting rule here exposed, which fixes a particular way for the voters to express their preferences and defines a clear algorithm for determining the final identification of the committee, could be implemented in practice. If our results are confirmed in such hypothetical real scenario, the algorithm discussed here will define an efficient form of democracy by delegation based on proxy voting \cite{boldi}, which robustly
 shows high level of representativeness and integrity of the selected committee.

\end{document}